\documentclass[% art
 aip, jcp, longer
 amsmath,amssymb, longbibliography,
preprint, %
 reprint,%
]{revtex4-1}

\usepackage{graphicx}% Include figure files
\usepackage{dcolumn}% Align table columns on decimal point
\usepackage{bm}% bold math
\usepackage{float}
\usepackage[utf8]{inputenc}
\usepackage[T1]{fontenc}\usepackage{amsmath}

\usepackage{comment}
\usepackage{amssymb}
\usepackage{xcolor}
\usepackage{float} %%need it to used [H] for figure placement
\usepackage{dsfont}

\begin{document}

%\title{Self-assembly limit cycles of signaling particles \vspace{10px}}
%\title{Emergent Limit Cycles in Self-Assembling  Signaling DNA-Coated Colloids}
%\title{Paracrine signaling induces cyclic colloidal self-assembly}
%\title{Paracrine signaling induces colloidal assembly-disassembly limit cycles}
%\title{Chemical signaling between colloids induces assembly-disassembly limit cycles}
%\title{Chemical signaling between colloids induces self-assembly limit cycles}
%\title{Chemical signaling drives colloidal self-assembly limit cycles}
\title{Synthetic paracrine signaling of colloids drives self-assembly limit cycles}
%automata, Brownian
%Titles of other people
%Non-reciprocal multifariousself-organization
%The hidden architecture of equilibrium self-assembly
%Pattern recognition in the nucleation kinetics of non-equilibrium self-assembly
%Temperature protocols to guide selective
%self-assembly of competing structures
%Multifarious assembly mixtures: Systems allowing
%retrieval of diverse stored structures

\author{Tim E. Veenstra}
\affiliation{Soft Condensed Matter \& Biophysics, Debye Institute for Nanomaterials Science, Utrecht University, Princetonplein 1, 3584 CC Utrecht, The Netherlands}
\author{Ren\'e van Roij}
\affiliation{Institute for Theoretical Physics, Utrecht University,  Princetonplein 5, 3584 CC Utrecht, The Netherlands}
\author{Pepijn G. Moerman}
\affiliation{Department of Chemical Engineering and Chemistry, Eindhoven University of Technology, 5612 AE Eindhoven, The Netherlands }
\author{Marjolein Dijkstra}
\affiliation{Soft Condensed Matter \& Biophysics, Debye Institute for Nanomaterials Science, Utrecht University, Princetonplein 1, 3584 CC Utrecht, The Netherlands}

\date{\today}
\begin{abstract} 
Developing synthetic materials that exhibit life-like behavior, such as internally driven cycles, remains a central challenge in active matter. Here, we introduce a minimal colloidal model of chemical signaling in which  particles produce   diffusing signaling molecules that selectively promote or inhibit attractive interactions among neighboring particles. This bio-inspired, paracrine-like signaling mechanism generates  context- and history-dependent many-body interactions that break time-reversal symmetry and drive the system far from equilibrium, leading to the spontaneous emergence of autonomous, internally sustained limit cycles in the composition of particle clusters. Using computer simulations, we map the resulting nonequilibrium phase behavior and identify distinct dynamical regimes controlled by the rates of signal production and degradation, together with the diffusion  range of the signaling molecules. Among these, we find a robust oscillatory state in which particle clusters autonomously assemble in a cyclic fashion, driven entirely by internal feedback loops. Our results establish 
paracrine-signaling colloids as a minimal, physically realizable platform for programmable nonequilibrium materials with life-like functionality and provide a general route toward synthetic active matter with self-regulated collective dynamics.

\end{abstract}
\maketitle

\section{Introduction}
The equilibrium self-assembly of colloidal particles into fluids, crystals, liquid crystals, quasicrystals, and other aggregated structures has been extensively studied \cite{Onsager1949, glotzer2007anisotropy, Sacanna2011, Manoharan2015, boles2016self, Li2020, Zhou2023, Gao2025}. The assembly is driven by interactions that arise from a wide range of mechanisms depending on particle properties such as shape~\cite{torquato2009dense,agarwal2011mesophase,de2011dense,damasceno2012predictive}, size~\cite{boles2016self,Barrat2023}, charge~\cite{leunissen2005ionic,shevchenko2006structural,hueckel2020ionic,mao2025stabilizing}, solvent environment~\cite{pieranski1980two,hertlein2008direct}, and thermodynamic state point. In recent decades, interactions mediated by biopolymers such as DNA have emerged as a particularly versatile tool for controlling colloidal assembly~\cite{Seeman2017,Jacobs2025}. These advances provide a platform for the programmable self-assembly of more complex structures \cite{Park2008, Nykypanchuk2008, McMullen2022, Moerman2026, Rogers2016, He2020, Melio2026}. DNA-coated colloids, for instance, have enabled the realization of a wide variety of synthetic materials, ranging from colloidal diamond structures~\cite{He2020} to colloidal foldamers~\cite{McMullen2022}, and colloidal metamaterials~\cite{Melio2026}.

A particularly important class of nonequilibrium phenomena involves the introduction of non-reciprocal interactions~\cite{Fruchart2021,Barrat2023,Dinelli2023}. In equilibrium, interactions must obey Newton’s third law; out-of-equilibrium systems, however, may break this action–reaction symmetry \cite{Ivlev2015,fruchart2026nonreciprocal}. Non-reciprocity leads to time-reversal symmetry breaking and enables the conversion of energy into useful work \cite{YLiu2024, JVeenstra2025, Brandenbourger2019, Saha2020, Rana2024, Huang2024, Osat2022, Osat2024, Metson2025, TEVeenstra2025}. One of the mechanisms through which non-reciprocity may be induced is catalytic activity, which can lead to predator-prey dynamics \cite{Meredith2020}. In this case, the generated chemical concentration gradients give rise to an asymmetric chemotactic force that acts either toward or away from the source.

 Another interesting direction in nonequilibrium dynamics is dissipative self-assembly, in which supramolecular structures are maintained through the continuous consumption of fuel~\cite{dellaSala2017}. These systems can reproduce certain characteristics of living matter~\cite{Tena-Solsona2018,DelGrosso2022,Hou2023}, relying on persistent energy consumption to sustain function and organization. Periodic assembly–disassembly cycles have, for instance, been realized by coupling the self-assembly of particles or nanoparticles to external chemical oscillators~\cite{Lagzi2010,VanRavensteijn2017,VanRavensteijn2020,Reja2024} or to driven reaction networks~\cite{Tena-Solsona2017,Dehne2019,Dehne2021,Sharma2023}.
However, dissipative self-assembly systems that exhibit internally generated limit cycles, where the driving mechanism emerges from the system itself rather than from  externally imposed signals, have so far remained elusive.

In living matter, interactions are rarely static; instead, dynamics can emerge from within the system itself. Biological agents, such as cells and organisms, actively produce signals that regulate both their own interactions and those of their neighbors. Cell–cell communication, for example, plays a central role in collective biological processes, such as morphogenesis, metastasis, and the formation of bacterial complexes~\cite{Su2024,Ellison2016,Bassler1999}. Cells employ multiple communication mechanisms \cite{Gilbert2000}. A well-known example is  endocrine signaling, where cells secrete hormones into the circulatory system to influence distant cells throughout the body. In contrast to this global endocrinical mode of communication, cells also interact locally through  \textit{paracrine} signaling, in which  signaling molecules are released that diffuse passively through the surrounding tissue. The range of these signals varies widely, from nearest neighbor communication \cite{Reilly1996, Jones1996}  to distances of approximately ten cell lengths \cite{Gurdon1994}. By coupling local interactions to dynamically evolving chemical environments, paracrine signaling endows tissues with adaptive and self-regulating behavior that is largely absent in conventional synthetic materials.

When designing synthetic out-of-equilibrium systems with life-like properties, colloidal systems provide a particularly promising platform due to the flexibility and tunability of their interactions \cite{Zeravcic2017a, Buerle2018,Grauer2024}. 
Moreover, complex behaviors such as self-replication and catalysis have been theoretically predicted to emerge in colloidal systems \cite{Zeravcic2014, Zeravcic2017b, Ouazan-Reboul2023}, suggesting the possibility of  mimicking biological  cycles, such as replication or metabolism. 
Experimentally, DNA-coated particles have recently been shown to possess binding sites whose interaction strength and specificity can be dynamically (re)programmed \cite{Moerman2023, Moerman2026, Xiong2024}. This capability becomes particularly powerful when combined with the production of RNA signals from DNA templates positioned on particle surfaces \cite{Dehne2019, Dehne2021, Kim2025}, making such systems a natural platform for developing signal-producing colloidal matter capable of emulating key features of paracrine communication. 

Here we introduce and explore an experimentally plausible colloidal model system of paracrine signal-producing particles and demonstrate that this paradigm enables autonomous self-assembly limit cycles. We show that long-term oscillations in the composition of colloidal clusters  emerge from the  interplay between signals that promote and inhibit particle  interactions. By mapping the phase behavior of the system, we identify distinct dynamical regimes governed by the properties of the signaling process. Finally, we show that asymmetric paracrine signaling provides a mechanism for breaking time-reversal symmetry, establishing a route to nonequilibrium behavior that is fundamentally distinct from conventional non-reciprocal interactions.

\section{Signal-producing DNA-coated colloids}

\begin{figure*}[t!]
    \centering
    \includegraphics[]{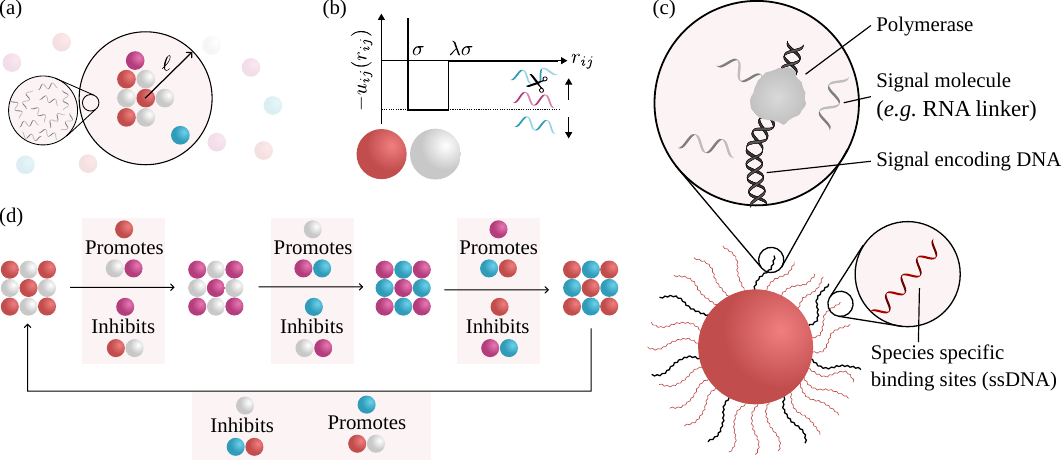}
    \caption{(a) Simplified model for the signal distribution around a particle. The signal concentration is assumed to be 
    nonzero and spatially constant within a  radius $\ell$ around the particle. (b) Schematic  interaction potential between two particles of different species. The depth of the attractive well is determined by the local signal concentrations that can either promote or inhibit binding. (c) Schematic illustration of a proposed mechanism for signal-producing colloids. Particles are functionalized with 
    signal-encoding double-stranded DNA, as well as single-stranded DNA (ssDNA) binding sites. The signal-encoding DNA can be transcribed by polymerases to produce signal molecules. (d) Graphical representation of the signaling rules defined in Eq.  (\ref{eq:policy_cyclic}) for $n=4$ species, where each species $\gamma$ promotes  binding between species $\gamma+1$ and $\gamma+2$, while  inhibiting  binding between  species $\gamma-1$ and $\gamma-2$. This leads to a cyclic sequence of programmed configurations.}
    \label{fig:model}
\end{figure*}

\begin{figure*}
    \centering
    \includegraphics{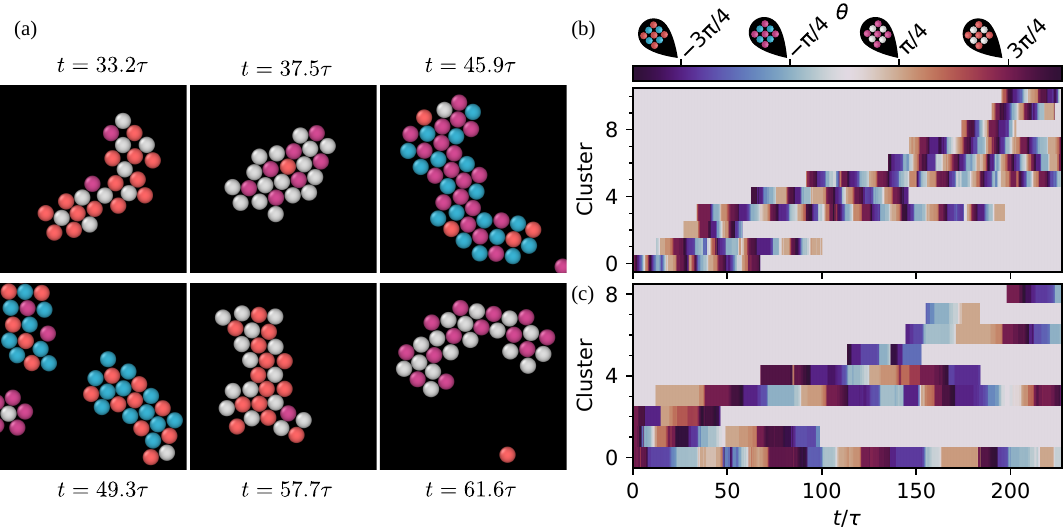}
    \caption{Typical configurations  showing self-assembly cycles. (a) Simulation snapshots for signal production rate $R_p\tau = 3.08$ and signal degradation rate $R_d\tau = 0.044$, and signal range $\ell/\sigma = 6$. One cluster is followed through time. (b-c) Kymographs of the phase $\theta_i$ of each cluster $i$ in the system as a function of time $t/\tau$ for  signal production rate $R_p \tau= 3.08$, with signal degradation rate (b) $R_d \tau= 0.22$ and (c) $R_d \tau= 0.044$.  }
    \label{fig:cyclical-assembly}
\end{figure*}
    
We propose a simple model of dynamical self-assembly driven by active signal production. This model describes the general case in which particles (or even cells) excrete diffusive signals that influence the binding-interactions of neighboring particles, see Fig. \ref{fig:model}(a-b). However, we also envision an experimentally realizable system of DNA-coated colloidal particles as shown in Fig. \ref{fig:model}(c). In this specific realization of this mechanism, each colloidal species carries species-specific binding sites composed of single-stranded DNA (ssDNA) that is covalently (irreversibly) bound to the colloidal surface. 
In addition, the colloidal particles are functionalized with double-stranded DNA that can be transcribed by polymerases (enzymes) to produce signaling molecules \cite{Dehne2019, Dehne2021, Kim2025}. This is an active, fuel-consuming (NTP-driven) process in which enzymes synthesize  RNA (anti)-linkers that modulate the binding strength between two colloidal species. For instance, RNA linkers can hybridize with the ssDNA grafted onto two particle species, thereby promoting their binding, whereas RNA anti-linkers can hybridize with the RNA linkers to inhibit particle binding. In this way, the interactions between particles depend on the signals produced on the surfaces of  neighboring particles in the recent past. Signaling  molecules are also enzymatically degraded in the bulk, ensuring a finite spatiotemporal range of the interaction memory \cite{Kim2026}. In this simulation study, we introduce a simplified model that captures these essential features of synthetic paracrine signaling on the basis of effective colloid-colloid interactions. 

We consider an $n$-component dispersion of $N$ colloidal spheres, which for simplicity all have the same diameter $\sigma$, but are chemically distinct. The system contains $N_\gamma$ particles of species $\gamma=1,2,\cdots,n$, such that $N=\sum_{\gamma=1}^nN_\gamma$. Each particle $i$ of species $\alpha$ interacts pairwise with particle $j$ of species $\beta$, separated by a center-to-center distance $r$, via a time-dependent square-well potential
\begin{equation}
    U_{ij}^{\alpha\beta}(r, t) = 
    \begin{cases} 
        \infty, &\text{if $r<\sigma$; }\\
        -u_{ij}^{\alpha\beta}(t) , &\text{if $\sigma\leq r \leq \lambda \sigma$}; \\ 
        0, &\text{if $r>\lambda\sigma$,}
    \end{cases}
    \label{eq:u}
\end{equation}
where $\lambda\sigma$ denotes the range of the (DNA-mediated) attraction and $u_{ij}^{\alpha\beta}(t) > 0$ is the time-dependent depth of the potential well between particles $i$ and $j$, see Fig.~\ref{fig:model}(b). The interaction range  $\lambda\sigma$ is assumed to be constant and independent of particle species, however the well depth  $u_{ij}^{\alpha\beta}(t)$ depends not only  on species $\alpha$ and $\beta$ but also on  time $t$ through the local chemical signal concentrations generated by nearby particles, their positions, and their recent environments. 
Chemical signals are produced locally by each colloid at a rate $R_p$, then diffuse with a diffusion coefficient $D$, and finally degrade on a timescale $\tau_d=1/R_d$, where $R_d$ is the spatially uniform degradation rate. These processes define an effective interaction length scale $\ell=\sqrt{D\tau_d}$, which characterizes the spatial extent of the emitted chemical signals around a signal-producing colloid, as illustrated schematically in Fig.~\ref{fig:model}(a). Rather than explicitly modeling the full spatiotemporal signal field, we  assume that the signal concentration is spatially uniform within a radius $\ell$ of the instantaneous center of mass of an isolated signal-producing colloid. Consequently, only particles located within this radius $\ell$ can be affected by the signal, which modifies their interaction strengths with one or more  colloid species. As a result, the well depth $u_{ij}^{\alpha\beta}(t)$ carries a memory of the recent local environments of the interacting particles $i$ and $j$.  In the SI, we show that replacing this approximation with a more realistic spatially varying signal concentration field yields qualitatively the same behavior \cite{SI}.

To capture the dependence of the binding strength on both the local environment and its recent history, we model the time evolution of  $u_{ij}^{\alpha\beta}(t)$ by 
\begin{align}
    \label{uij}
    \hspace{3mm}\frac{\partial u^{\alpha\beta}_{ij}}{\partial t} =& -R_d u^{\alpha\beta}_{ij} + \\ &\hspace{-12mm} R_p k_BT\sum_{\gamma=1}^{n}\sum_{k=1}^{N_\gamma} \frac{1}{2}\Big(H(\ell-r_{ik})+H(\ell-r_{jk})\Big)s_{\alpha\beta\gamma}. \notag 
\end{align}
Here we recall that $R_p$ and $R_d$ are the signal production and degradation rates per colloid, respectively, and  $N_\gamma$ is the number of colloids of species $\gamma\in\{1,2,\cdots,n\}$ in this $n$-component mixture. The first term in Eq. (\ref{uij}) describes the decay of the interaction strength due to signal degradation, while the second term accounts for signal production by nearby particles. Signal-induced changes of the well depths, which promote or inhibit binding, can only be triggered by  particles within the signal range $\ell$, as enforced by the Heaviside step function $H(\ell-r)$ in Eq. (\ref{uij}). Thus,   
all particles $k$ of species $\gamma$ that lie within a distance $\ell$ of either particle $i$ or $j$ of species $\alpha$ and $\beta$, respectively, contribute to the rate of change of the interaction strength $u_{ij}^{\alpha\beta}$. The pair potential  in Eq.(\ref{eq:u}) therefore represents an effective many-body interaction. The sign and magnitude of these contributions are determined by the dimensionless rank-three matrix $s_{\alpha \beta \gamma}$, whose elements are typically of order unity (or zero). The characteristic energy contribution per production event is  of order  $k_BT$, with $k_B$ the Boltzmann constant and $T$ the temperature. 
The precise form of $s_{\alpha \beta \gamma}$ determines the specific signal sequences produced by the different colloid species, where positive (negative) values give rise to an increase (decrease) of the well depth of the $\alpha\beta$ interaction by nearby particles of species $\gamma$. 

\section{Self-assembly limit cycles}
The paracrine-signaling framework introduced here can, in principle, generate a broad range of static and dynamic nonequilibrium phenomena, including self-limiting growth, predator–prey dynamics, heterogeneous catalysis, and onion-shell self-assembly, through appropriate choices of the number of particle species, the signal production and degradation rates, and the signal-encoding rank-three tensor, see SI~\cite{SI}. Here, we focus on its application to the design of self-assembly limit cycles of binary structures. 
To be specific, we consider the case $n=4$  with colloidal species $\gamma\in\{1,2,3,4\}$. Starting from a binary cluster composed of species 1 and 2 (shown in red and white in Fig.~\ref{fig:model}(d)), while species 3 and 4 remain dispersed throughout the system, the cluster transitions into a binary 2-3 cluster due to a judicious design that weakens 1-2 and strengthens 2-3 attractions, followed likewise by a transition to a 3-4 and 4-1 cluster, before returning to the initial 1-2 cluster, thereby forming a closed cycle. Such a sequence can be realized if the signal produced by species $\gamma$ weakens the attractive interaction $u_{\alpha\beta}$ for $(\alpha,\beta)=(\gamma-2,\gamma-1)$ and strengthens it for $(\alpha,\beta)=(\gamma+1,\gamma+2)$, where $\gamma\pm1$ and $\gamma\pm2$ are understood to be modulo 4. The paracrine signaling is therefore inherently asymmetric: a given species suppresses the formation of clusters that occur earlier in the cycle while promoting the formation of clusters that occur later, thereby imposing a preferred direction on the cyclic dynamics. 
For $n=4$, these combined promotion-inhibition signaling rules  are mathematically encoded  by
\begin{align}
    s_{\alpha\beta\gamma} = & -\delta_{\alpha, \beta-1} \left(  \delta_{\alpha, \gamma-2} - \delta_{\alpha, \gamma+1} \right) \notag \\ 
    & -  \delta_{\alpha,\beta+1} \left( \delta_{\alpha, \gamma-1} - \delta_{\alpha, \gamma+2} \right),
    \label{eq:policy_cyclic}
\end{align}
where all indices are taken modulo $4$ (and where we do \emph{not} use the Einstein convention for summation). For simplicity, we assume in Eq.(\ref{eq:policy_cyclic}) that all signaling contributions have the same magnitude. One checks that $s_{\alpha\beta\gamma}=s_{\beta\alpha\gamma}$, such that the direct interaction potential is reciprocal, $U^{\alpha\beta}_{ij}=U_{ji}^{\beta\alpha}$. For convenience, a matrix representation of $s_{\alpha\beta\gamma}$ for $\gamma\in\{1,2,3,4\}$ is given in the Methods section. A possible physical realization consists of four colloidal species, labeled $\gamma$, each carrying DNA templates that produce two distinct signaling molecules: one that inhibits binding between species $\gamma-2$ and $\gamma-1$, and another that promotes  binding between species $\gamma+1$ and $\gamma+2$.  

We perform Monte Carlo simulations of an equimolar four-component mixture of $N=72$ particles (so $N_\gamma=18$ for $\gamma=1,2,3,4$) of diameter $\sigma$ in a periodic two-dimensional simulation box such that the total packing fraction equals $\eta = 0.025$. The time scale $\tau$ for colloids to diffuse over the typical interparticle spacing is taken as our unit of time (see Methods for details). 
Using the rank-three matrix  $s_{\alpha\beta\gamma}$ defined in Eq. (\ref{eq:policy_cyclic}), we observe  sustained self-assembly cycles for a signal production rate $R_p\tau =3.08$, signal degradation rate $R_d\tau=0.044$, and signal range $\ell/\sigma=6$. 

\begin{figure*}
    \centering
    \includegraphics[]{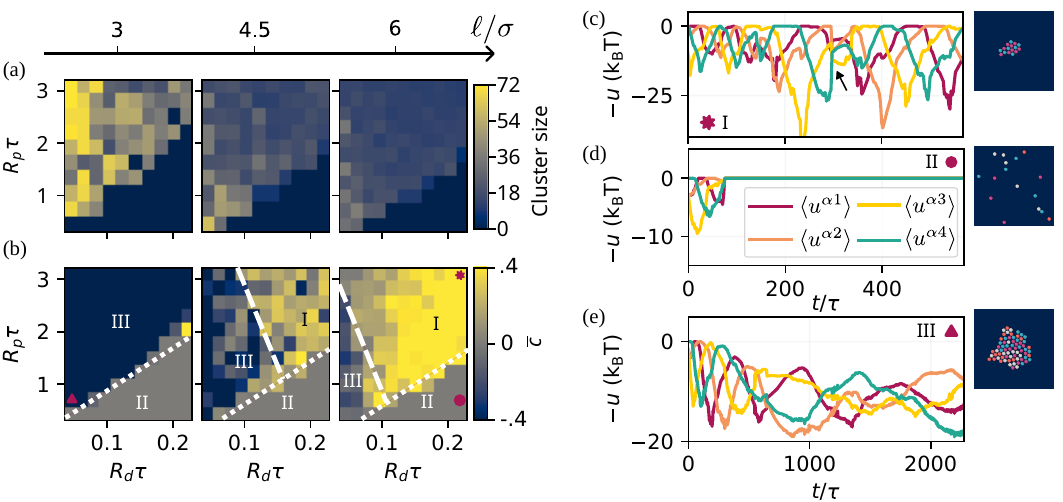}
    \caption{(a) Average cluster size and (b) average  anti-correlation parameter $\bar{c}$ as a function of signal production rate $R_p \tau$ and signal degradation rate $R_d \tau$, for three different signal radii $\ell/\sigma=3, 4.5,$ and 6. Three distinct regimes can be identified:  Regime I corresponds to self-assembly limit cycles, regime II to melting, and regime III to uncontrolled aggregation. Dashed and dotted lines  are guides to the eye, indicating the approximate boundaries between the regimes. (c-e) The average well depth $\langle u^{\alpha\beta}\rangle$ of particles of species $\alpha$ and $\beta$ within a single cluster (the initialized cluster) with all four possible species $\beta$, as a function of time, for (c) $R_p\tau = 3.08$, $R_d\tau=0.22$,  $\ell/\sigma=6$; (d)  $R_p\tau = 0.88$, $R_d\tau=0.22$,  $\ell/\sigma=6$; and (e)  $R_p\tau = 0.88$, $R_d\tau=0.044$,  $\ell/\sigma=3$. The arrow in (c) marks a merging event. The insets show representative configurations of the tracked cluster during the simulations.  }
    \label{fig:fig3}
\end{figure*}

This is illustrated in Fig. \ref{fig:cyclical-assembly}(a), where we show snapshots of a system that has been initialized such that particles of two species (say 1 and 2, represented by red and white) form a binary 1-2 crystal with an initial mutual interaction strength of $7 {\rm~k_BT}$, while the other two species (say 3 and 4, represented by purple and blue) are homogeneously dispersed throughout the simulation box. Starting from this configuration, the system indeed sequentially  transitions through all programmed 2-3, 3-4, and 4-1 crystal structures shown in Fig. \ref{fig:model}(d), before eventually returning to the  initial 1-2 crystal configuration, after which the cycle starts again in a sustained fashion. Importantly, this behavior does not depend on the initial condition as cyclically evolving clusters  also spontaneously nucleate from a homogeneous fluid, as can be seen in the supplemental videos. In that case, however, multiple distinct clusters coexist and independently progress through the same assembly cycle.

We characterize the cyclic behavior of the $i$-th  cluster using a phase angle $\theta_i\in[0,2\pi]$, which  specifies the phase of  cluster $i$  along the assembly cycle in phase space  (see Methods and Fig.~\ref{fig:model}(d)). In Figs. \ref{fig:cyclical-assembly}(b-c), we plot $\theta_i$  as a function of time for all clusters in two simulations with signal production rate $R_p\tau=3.08$ and degradation rates (b) $R_d\tau = 0.22$ and (c) 0.044. In both cases, robust self-assembly cycles emerge, with all clusters repeatedly traversing the full sequence of states. The oscillation period, however, decreases with increasing degradation rate.
Clusters are continuously created and destroyed through nucleation, dissolution, coalescence, and fragmentation. As shown in Figs. \ref{fig:cyclical-assembly}(b-c), newly formed clusters also exhibit the same cyclic dynamics, irrespective of their initial phase.

We next explore the region in parameter space in which this behavior occurs. In Fig. \ref{fig:fig3}(a), we show the average cluster size as a function of the signal production rate $R_p\tau$ and degradation rate $R_d \tau$ for different signal radii $\ell/\sigma=3, 4.5$, and 6. For simplicity, we treat $\ell$ and $R_d$ as independent parameters. At low values of $\ell$ and  $R_d$, we observe the formation of very large clusters, indicative of  uncontrolled aggregation of all species. In contrast, at high degradation rates and low production rates, we observe melting of all clusters and the system remains dispersed. This behavior reflects the fact that the assembled structures in our model are not thermodynamically stable  equilibrium states. Instead, their persistence relies on sustained signal production, which drives the transition to the next structure in the cycle before the current one dissolves as the signals degrade. When signal production is too weak or signal degradation is too rapid, this balance is lost, the cycle breaks down, and the structures disassemble.

To quantify the extent to which  clusters follow the cycle shown in  Fig. \ref{fig:model}(d), we introduce for each cluster $i$ an anti-correlation parameter $c_i$ between species  $\alpha$ and $\alpha+2$ (mod $4$), averaged over  all species $\alpha$ (see Methods). By construction, $c_i$ is positive when the fractions of species $\alpha$ and $\alpha+2$ within a  cluster are anti-correlated, corresponding to self-assembly cycles; negative when they are correlated, indicative of uncontrolled aggregation; and zero in the absence of any correlation. In Fig. \ref{fig:fig3}(b), we show the average value $\bar{c}$ of this cyclic order parameter as a function of the production rate $R_p\tau$ and degradation rate $R_d\tau$ for signal radii $\ell=3, 4.5,$ and 6, and where we define $\bar{c}$ to be the average over all clusters and time steps of all simulations performed at a fixed $R_p$ and $R_d$.  Three  qualitatively different regimes can be identified. At high degradation rates, the cyclic order parameter $\bar{c}$ is close to zero (regime II in Fig. \ref{fig:fig3}(b)). In this regime, interparticle attractions are too weak to sustain stable assemblies, resulting in small clusters that readily dissolve. At low degradation rates and high production rates (regime III in Fig. \ref{fig:fig3}(b)), the anti-correlation parameter $\bar{c}$ becomes negative, indicating that species $\alpha$ and $\alpha+2$ tend to be correlated within the same cluster. In this regime,  the system forms large aggregates containing all species, without any compositional ordering corresponding to the programmed cycle. For sufficiently large $\ell$, however, we can identify a broad intermediate dynamic regime of production and degradation rates in which clusters exhibit strong anti-correlations between species $\alpha$ and $\alpha + 2$, corresponding to positive values of $\bar{c}$. This regime is indicative of robust cycles of self-assembly (regime I).

\section{Asymmetric paracrine interactions}
To gain insight into the mechanism underlying the different  dynamical behaviors, we  examine how the signal-mediated interactions evolve in time. We monitor for each species $\beta=1,2,3,4$ the time-dependent  average well depth $\langle u^{\alpha\beta}\rangle$, where the average $\langle \ldots\rangle$ is taken over the particles of species $\alpha\neq\beta$ within the initialized cluster.

In Fig. \ref{fig:fig3}(c), we show the  average well depth at a signal production rate $R_p\tau=3.08$, degradation rate $R_d\tau=0.22$, and signal radius $\ell/\sigma=6$, which lies well within the cyclic regime I (as indicated by the red star in Fig. \ref{fig:fig3}(b)). We observe in Fig. \ref{fig:fig3}(c) that the interaction strengths associated with  the four species oscillate in time and return to $\langle u^{\alpha\beta}\rangle=0$ at the end of each cycle. Furthermore, we find that the limit cycle is robust to  merging events, seen at approximately  $t=300\tau$ (indicated with an arrow), where the average interaction strength of all species exhibits a sharp increase, after which the system relaxes back to the regular oscillatory dynamics. 

In contrast, Fig. \ref{fig:fig3}(d) shows the interactions of a cluster in the melting regime II, at a signal production rate $R_p\tau=0.88$, degradation rate $R_d\tau=0.22$, and signal radius $\ell/\sigma=6$ (as indicated by the red dot in Fig. \ref{fig:fig3}(b)). Here,  the extrema in the interaction strength gradually decrease over time until they fall below a threshold, at which point the cluster completely dissolves. Finally, the curves in Fig. \ref{fig:fig3}(e) correspond to the aggregated regime III, at a signal production rate $R_p\tau=0.88$, degradation rate $R_d\tau=0.044$, and signal radius $\ell/\sigma=3$ (as indicated by the red triangle in Fig. \ref{fig:fig3}(b)). Here,  interaction energies do not fully decay before the onset of the next cycle, leading to the gradual accumulation of bonds between all species. As a result,  all binary structures shown in Fig. \ref{fig:fig3}(e) coexist within a single aggregate. However, none of  these configurations is individually stable. Instead, self-assembly cycles persist only locally, and the composition oscillations become desynchronized across the entire cluster. 
Moreover, the oscillation period  is governed primarily  by the signal degradation rate rather than the signal production rate, resulting in dynamics that is significantly slower than those observed in the regime of self-assembly limit cycles.

\begin{comment}
\begin{figure*}[t!!]
    \centering
    \vspace{5px}
    \includegraphics[]{NewFigures/Fig4.pdf}
    \caption{(a-b) Average interaction strength $-\langle u^{\alpha,\alpha+1}\rangle/k_BT$ as a function of (red) the number of neighbors $n_{\alpha+2}$ of species $n+2$, and (orange) the number of neighbors $n_{\alpha-1}$ of species $\alpha-1$. Data was obtained at signal production rate $R_p\tau = 3.08$, degradation rate $R_d\tau=0.22$ and signal range $\ell/\sigma=6$. These simulations were performed for (a) the regular system with both promotion and inhibition of interactions, as described in Eq. \ref{eq:policy_cyclic}, and (b) a modified system in which  particles can only promote interactions, while inhibition is only mediated globally through signal degradation. (c) The  anti-correlation parameter $\bar{c}$, as defined in Eq.  (\ref{eq:correlations}), in the uninhibited system as a function of production rate $R_p\tau$ and degradation rate $R_d\tau$. (d-e) Typical configurations  in the uninhibited system at $R_p\tau = 3.08$, $R_d\tau=0.22$, and (d) $\ell/\sigma=6$ and (e) $\ell/\sigma=3$. Error bars represent the standard deviation of the mean.
    }
    \label{fig:NR}
\end{figure*}
\end{comment}

\begin{figure}
    \centering
    \includegraphics[width=\linewidth]{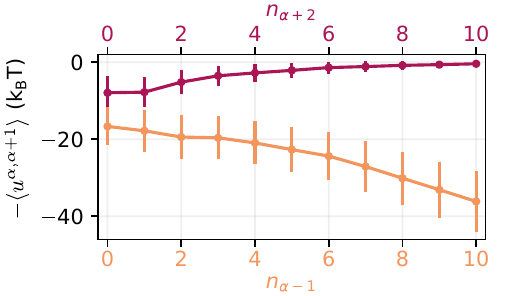}
    \caption{Average interaction strength $-\langle u^{\alpha,\alpha+1}\rangle/k_BT$ as a function of (red) the number of neighbors $n_{\alpha+2}$ of species $n+2$, and (orange) the number of neighbors $n_{\alpha-1}$ of species $\alpha-1$. Data was obtained at signal production rate $R_p\tau = 3.08$, degradation rate $R_d\tau=0.22$ and signal range $\ell/\sigma=6$. Error bars represent the standard deviation of the mean.}
    \label{fig:NR}
\end{figure}

We emphasize that all pairwise interactions in our model for paracrine signaling are explicitly reciprocal; every bond obeys $\langle u^{\alpha,\alpha+1}\rangle = \langle u^{\alpha+1,\alpha}\rangle$ at all times, so the directionality in the cyclic  processes results from a mechanism that fundamentally differs from the two-body non-reciprocal interactions presented in e.g.  Refs.~\citenum{Soto2014self} and \citenum{Meredith2020}. 
In our model the time-reversal symmetry breaking stems from a many-body effect: the interaction between species $\alpha$ and $\alpha +1$ depends on the presence of species $\alpha +2$ in a different way than  on the presence of species $\alpha-1$.
This effect is illustrated in Fig. \ref{fig:NR} where we compare  the average interaction strength $\langle u^{\alpha,\alpha+1}\rangle$ between species $\alpha$  and $\alpha+1$ as a function of the numbers $n_{\alpha-1}$ and $n_{\alpha+2}$ of neighboring particles of species $\alpha-1$ (orange) and species $\alpha+2$ (red), respectively. 
As $n_{\alpha-1}$ increases, $-\langle u^{\alpha,\alpha+1}\rangle$ becomes increasingly negative, indicating stronger attraction. By contrast, the interaction weakens with increasing $n_{\alpha+2}$ and even approaches zero for sufficiently large $n_{\alpha+2}$. Interestingly, the two interaction strengths are not necessarily equal at zero neighbors, when $n_{\alpha-1}=n_{\alpha+2}=0$, reflecting a history dependence of the interactions. Particles of species $\alpha$ that are currently in a binary structure with species $\alpha+1$ have, on average, spent more time in the vicinity of species $\alpha-1$ than particles of species $\alpha+1$ have spent near species $\alpha+2$. As a result, the signal-induced enhancement of the interaction strength $\langle u^{\alpha,\alpha+1}\rangle$ persists even after the relevant neighbors are no longer present.

The directionality in our model therefore does not originate from non-reciprocal pairwise forces, but instead emerges from the production of paracrine signals. The effective interaction between a given pair of colloids depends not only on their instantaneous separation, but also on their evolving local environment and the system's history through the time-dependent interaction well depth as described in Eq. (\ref{uij}). The resulting cyclic self-assembly dynamics is therefore a genuinely emergent many-body phenomenon driven by asymmetric paracrine signaling, rather than a consequence of intrinsic pairwise non-reciprocity.  The cyclic behavior is encoded in the  signaling policy defined in Eq. (3), which combines local signal-mediated promotion ($s_{\alpha\beta\gamma}>0$) of specific interactions with inhibition ($s_{\alpha\beta\gamma}<0$) of others. In the SI, we demonstrate that interaction-inhibiting signals are crucial as their  omission effectively eliminates cyclic behavior.   

\begin{figure*}[t!]
    \centering
    \includegraphics[]{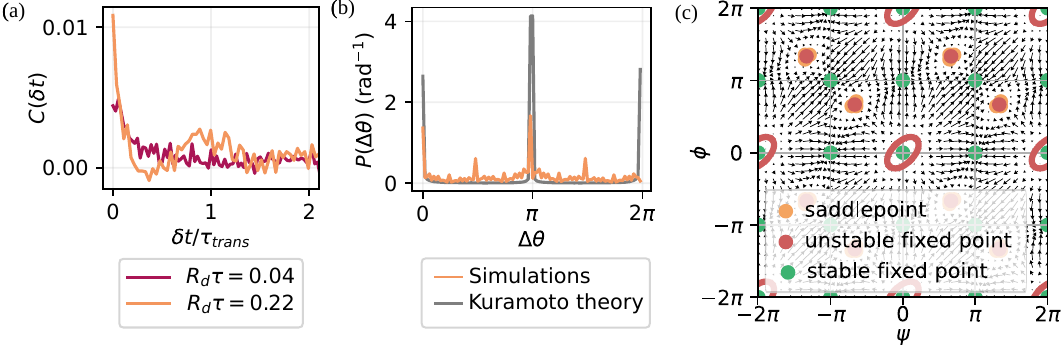}
    \caption{Dynamical behavior of clusters. (a) The time correlation  $C(\delta t)$ between transitions of two different clusters as a function of delay time $\delta t/\tau_{trans}$. (b) The probability distribution $P(\Delta\theta)$ of relative phase differences  between clusters at $R_p\tau=3.08$, $R_d\tau=0.22$, and $\ell/\sigma=6$, together with  results from a theoretical model  of  coupled Kuramoto oscillators. (c) Vector fields $(\partial_t \psi, \partial_t \phi)$ of the Kuramoto model as a function of phase differences $\psi$ and $\phi$. The points and lines correspond to fixed points of various stability. }
    \label{fig:fig5}
\end{figure*}

\section{Cluster dynamics}

We further investigate the dynamics of clusters in the cyclic regime I.  We first determine the transition time $\tau_{trans}$, which is the characteristic time scale associated with transitions between successive structures in the cycle, see Methods. Dense clusters in the limit-cycle regime coexist with a dilute surrounding dispersion with a composition determined by the clusters within it. 
As a result, the availability of species that are required to progress through  the self-assembly cycle is 
constrained by the phase of all nearby  clusters. This suggests the emergence of an effective, fluid-mediated interaction between clusters  that directly influences their  transition rate \cite{Toiya2010}. We investigate this interaction by computing the time correlation $C(\delta t)$ between transitions of two different clusters (see Methods) with a delay time $\delta t$. The results are shown in Fig. \ref{fig:fig5}(a) for a high production rate $R_p\tau=3.08$ at two different degradation rates $R_d\tau=0.04$ (red) and 0.22 (orange), respectively, as a function of the delay time $\delta t$ expressed in units of the transition time $\tau_{trans}$. We observe a clear correlation at both degradation rates, which is more pronounced at the higher $R_d$, indicating  that a transition of one cluster increases the  probability of triggering a (later) transition in another  cluster. Furthermore, for $R_d\tau=0.22$, the transitions are sufficiently regular that a secondary  peak emerges in the correlation function, reflecting the periodic nature of the underlying dynamics. 

However, clusters are not only correlated in time through synchronized transitions but also their instantaneous phase angles $\theta$ within the cycle are correlated. 

To quantify this correlation, we compute from our simulations the equal-time probability distribution $P(\Delta\theta)$ of the relative phase differences $\Delta\theta \equiv \theta_i-\theta_j$ between clusters $i$ and $j$. For $R_d\tau=0.22$, with all other parameters identical to those  in Fig.\ref{fig:fig5}(a), we show $P(\Delta\theta)$ by the orange curve in Fig.\ref{fig:fig5}(b). Based on symmetry, one might expect peaks of equal weight at all phase differences $\Delta\theta=n\pi/4$, where   $n\in \{0,1,2,3\}$. Instead, pronounced peaks appear only at  $\Delta\theta=0$ and $\Delta \theta=\pi$, whereas the peaks at $\Delta\theta=\pi/2$ and $\Delta\theta=3\pi/4$ are strongly suppressed. This indicates that the clusters spontaneously partition into  two groups that are internally (anti-)phase-synchronized but oscillate with a relative phase shift of $\pi$. The clusters formed through this paracrine signaling framework can therefore be considered coupled oscillators that exhibit both synchronization and anti-synchronization.  

We qualitatively reproduce this spontaneous symmetry breaking (gray curve in Fig.\ref{fig:fig5}(b)) using a theoretical model of three coupled non-Brownian Kuramoto oscillators with identical eigenfrequencies and pairwise coupling, as described in the SI \cite{SI}. We visualize the attractor states of this model in Fig. \ref{fig:fig5}(c) for the same parameter set as the gray line in Fig. \ref{fig:fig5}(b). The phase differences $\phi$ and $\psi$, defined relative to a reference oscillator, exhibit stable fixed points precisely at the states where two oscillators are synchronized and the third is anti-synchronized, namely $(\psi, \phi) = (0, \pi), (\pi,0), (\pi,\pi)$. Another stable fixed point occurs at $(0,0)$,  corresponding to complete synchronization of all three oscillators. However, this fixed point has only a small basin of attraction and is separated from the remainder of phase space by a non-isolated unstable fixed point. Consequently, the vast majority of trajectories evolve toward  the (anti)-synchronization partitioning, consistent with the behavior observed in our simulations of paracrine-signaling particles.

\section{Conclusions}
In conclusion, we have designed an experimentally realizable four-component colloidal system in which paracrine-like chemical  signaling drives autonomous self-assembly limit cycles  in a robust parameter space. In particular, the system undergoes a cyclic sequence of binary structures, $1-2 \rightarrow 2-3 \rightarrow 3-4 \rightarrow 4-1 \rightarrow  1-2$, thereby breaking time-reversal symmetry. This behavior 
arises from colloid-mediated signaling that selectively promotes and inhibits attractive interactions between neighboring particles,  giving rise to history- and environment-dependent effective interactions. Despite this nonequilibrium behavior and time-reversal symmetry breaking, the microscopic interactions remain reciprocal at all times and therefore satisfy Newton's third law.

An extensive investigation of  qualitatively distinct dynamic behaviors reveals three parameter regimes for signal production and degradation rates. We identify a broad regime (I) in which limit cycles of self-assembling clusters is remarkably robust. At lower degradation rates (regime III), we observe large aggregates that  still exhibit local cyclic dynamics  but lack global phase 
synchronization. At the sufficiently low production rates of regime II, however, the system melts entirely and forms a dilute gas-like state. This underscores the intrinsically non-equilibrium nature of these clusters,  which are unstable in equilibrium (i.e. without chemical signaling) and require a sufficiently but not overly strong signal production and degradation to sustain their  dynamic assembly.

More generally, paracrine signal-producing particles represent a design principle rather than a specific system. Any particle system in which interactions depend on signals emitted by nearby particles can, in principle, exhibit a wide range of collective behaviors,  including tunable morphology, composition, and dynamics. As such, these systems may  provide a minimal physical model for more complex biological signaling networks. With concrete experimental realizations becoming increasingly feasible, signal-producing particles offer a route toward programmable assembly and disassembly.

\section*{Acknowledgments} 
TEV and MD acknowledge  funding from the European Research Council (ERC) under the European Union's Horizon 2020 research and innovation programme (Grant agreement No.\ ERC-2019-ADG 884902 SoftML). PGM acknowledges funding from NWO through VI.Veni.222.194.

\section*{Contributions}
Conceptualization: TEV, RvR, PGM, MD. 
Funding acquisition: MD.
Investigation: TEV.
Methodology: TEV, RvR, PGM, MD.
Supervision: RvR, MD.
Writing – original draft: TEV.
Writing – review and editing: all authors.

\section*{Competing Interests Statement} 
The authors declare no competing interests.

\bibliography{library}

\vspace{10px}
\section{Methods}
\subsection{Simulation methods}
We perform Monte Carlo (MC) simulations of a two-dimensional equimolar four-component system described by Eqs.(\ref{eq:u}), (\ref{uij}), and (\ref{eq:policy_cyclic}) for a range of signal production rates $R_p$, signal degradation rates $R_d$,  and signal ranges $\ell$. The use of a MC scheme is  justified by the separation of timescales between particle self-assembly and (slow) transitions between clusters.  For colloidal hard-core diameter $\sigma$ of all species, we set the range of the colloidal attractions to $\lambda\sigma=1.05\sigma$, and constrain the interaction strengths $-u_{ij}^{\alpha\beta}$ to be strictly negative. We consider a relatively low packing fraction of $\eta=0.025$, such that the typical distance between colloidal particles exceeds the signal range $\ell < \sigma\eta^{-1/2}\simeq 6.32\sigma$ for all $\ell$ used in this work. As a result, the signal concentration  experienced by the freely dispersed colloids is negligible throughout. 

To establish a physical timescale for the MC simulations, we measure the mean-squared displacement of the colloids at low packing fraction and extract the corresponding effective diffusion coefficient $D_c$ of the colloids. We obtain $D_c = 3.6 \cdot 10^{-4} {\rm \sigma^2/\tau_{MC}}$, where $\tau_{MC}$ denotes a single MC time step. We then express all  timescales in terms of  the diffusive timescale $\tau$, which is defined as 
\begin{equation}
    \tau = \frac{4\sigma^2}{D_c\eta}, 
\end{equation}
where $\eta/4$ is the packing fraction of each colloidal species in the equimolar mixture. This timescale $\tau$ corresponds to the diffusion time required for a colloid to travel the typical distance needed to replace a particle within a cluster and therefore depends on the packing fraction.  For $\eta=0.025$ and the measured value of $D_c$, we find $\tau=4.4\cdot 10^5 \tau_{MC}$, which is the unit of time  used throughout this study. \vspace{5mm}

\subsection{Order parameters}

We quantify the phase angle $\theta_i$ of cluster $i$ within its cycle as 
\begin{equation}
    \theta_i = \text{arctan2}\left( \frac{\phi_1^i+\phi_2^i - \phi_3^i-\phi_4^i}{\phi_1^i-\phi_2^i-\phi_3^i +\phi_4^i}\right),
\end{equation}
where $\phi_a^i$ denotes the fraction of bonds in  cluster $i$ that corresponds to structure $a$ in the cycle, as defined in Fig. \ref{fig:model}(d), and where $\text{arctan2}(y/x)$ is the two-argument \mbox{arctangent} that is $2\pi$-periodic and returns the correct angle for each quadrant of the $(x,y)$-plane. The binary clusters are therefore with equal probability positioned on the unit circle at angles $-3\pi/4$, $-\pi/4$, $\pi/4$ and $3\pi/4$, as indicated in Fig. \ref{fig:cyclical-assembly}(b).

As also mentioned in the main text, we consider the anti-correlation parameter $c_i$ for cluster $i$, which is defined as
\begin{equation}
    c_i =   \sum_{a=1}^{4} \left(\langle\phi^i_a\rangle_t \langle\phi_{a+2}^i\rangle_t -  5 \langle\phi_a^i\phi_{a+2}^i\rangle_t  \right),
    \label{eq:correlations}
\end{equation}
where $\sum_{a=1}^{4}$ sums over all four binary structures in the system.  
Here, $\phi_a^i$ is again the fraction of bonds in cluster $i$ corresponding to binary structure $a$, and $\langle\dots\rangle_t$ is an average over the lifetime of this cluster.

In Fig. \ref{fig:fig3}, we calculate the average strength of the attractive interactions between particles in a certain cluster. For a given species $\beta$ we define this average bond energy as 
\begin{equation}
    \langle u^{\alpha \beta}\rangle = \sum_{\alpha=1}^n \sum_{j=1}^{N_\beta}  \sum_{\substack{i=1}}^{N^{cl}_\alpha} \frac{u_{ij}^{\alpha\beta}}{nN_\beta N^{cl}_\alpha}, 
\end{equation}
where we sum over all $N_\beta$ particles of species $\beta,$ all $n$ species $\alpha$, and all $N_\alpha^{cl}$ particles of species $\alpha$ that are within the  considered cluster.

In Fig. \ref{fig:fig5}(a), we rescale time with a characteristic timescale associated with transitions between successive  structures in the cycle. This timescale is obtained from the Fourier transform of $\sin{\theta_i(t)}$ for each cluster 
$i$. For each cluster, we identify the dominant frequency and then average this frequency over all clusters. The first 100 time steps are discarded to ensure that the system has reached a steady state. This resulting frequency corresponds to the entire self-assembly cycle, which we then convert into the transition time by dividing the period by four.

Subsequently, in the main text, we analyze the correlation $C(\delta t)$ between transitions of different clusters as a function of lag time $\delta t$. This quantity is defined as
\begin{align}
    C(\delta t) =  \sum_{t_i=0}^{N_t}& \frac{\tau_{trans}}{\Delta t ~N_c(t_i)N_c(t_i+\delta t)} \\ &  \times \sum_{c_1=1} ^{N_c(t_i)} \sum_{\substack{c_2=1 \\ c_2 \neq c_1}}^{N_c(t_i+\delta t)}  T_{c_1
    }(t_i)T_{c_2}(t_i+\delta t) \notag,
\end{align}
where $N_c(t)$ denotes the number of clusters in the system at time $t$, and the inner sum runs over all pairs of distinct clusters. The first summation runs over all $N_t$ sampled time steps $t_i$, separated by  intervals $\Delta t$. The transition signal is defined as the discrete time derivative of the phase with the mean drift removed, $T_{c}(t) = \theta_c(t_{i+1}) - \theta_c(t_{i}) - \Delta t\langle \dot\theta \rangle$, where $\langle\dot\theta\rangle$ is the average phase velocity over all clusters and times.

\subsection{Signal sequence rank-three matrices}

The specific signaling policy $s_{\alpha\beta\gamma}$ defined in  Eqs. (\ref{eq:policy_cyclic}) of the main text where it is represented as a rank-three matrix, can also be represented as a set of rank-two matrices. We can represent the influence of the signal from a single species, \textit{e.g.} species 1, as a regular rank-two matrix $s_{\alpha\beta 1}$. This represents the change in interactions between species $\alpha$ and $\beta$ due to the presence of species $1$. We can do this for all four species, which yields
\begin{widetext}
\begin{align*}
    s_{\alpha\beta 1} &= \left(
    \begin{matrix} 
        0 & 0 & 0 & 0 \\
        0 & 0 & 1 & 0 \\
        0 & 1 & 0 & -1 \\
        0 & 0 & -1 & 0 \\
    \end{matrix}\right)_{\alpha\beta} && 
    s_{\alpha\beta 2} = \left(
    \begin{matrix} 
        0 & 0 & 0 & -1 \\
        0 & 0 & 0 & 0 \\
        0 & 0 & 0 & 1 \\
        -1 & 0 & 1 & 0 \\
    \end{matrix}\right)_{\alpha\beta} && 
    s_{\alpha\beta 3} = \left(
    \begin{matrix} 
        0 & -1 & 0 & 1 \\
        -1 & 0 & 0 & 0 \\
        0 & 0 & 0 & 0 \\
        1 & 0 & 0 & 0 \\
    \end{matrix}\right)_{\alpha\beta} && 
    s_{\alpha\beta 4} = \left(
    \begin{matrix} 
        0 & 1 & 0 & 0 \\
        1 & 0 & -1 & 0 \\
        0 & -1 & 0 & 0 \\
        0 & 0 & 0 & 0 \\
    \end{matrix}\right)_{\alpha\beta}. 
\end{align*} 
\end{widetext}
We note that the sum of two consecutive matrices \textit{e.g.} $s_{\alpha\beta 0} + s_{\alpha\beta 1}$ represents the net signal generated by a binary 
cluster. In this case, the promotion signal produced by one species is canceled by the inhibitory signal produced by  its binary partner. As a result, only the interactions corresponding to  the next structure in the cycle are promoted, while those corresponding to  the previous structure are inhibited. This representation also shows that aggregates that contain all species are unstable, as discussed in the main text. Since the sum of all 4 matrices equals the null-matrix, the net signal produced by such an aggregate is zero, preventing the stabilization of the  aggregated state. 

\setcounter{figure}{0} 
\renewcommand{\figurename}{Extended Fig.}
\begin{figure*}
    \includegraphics[]{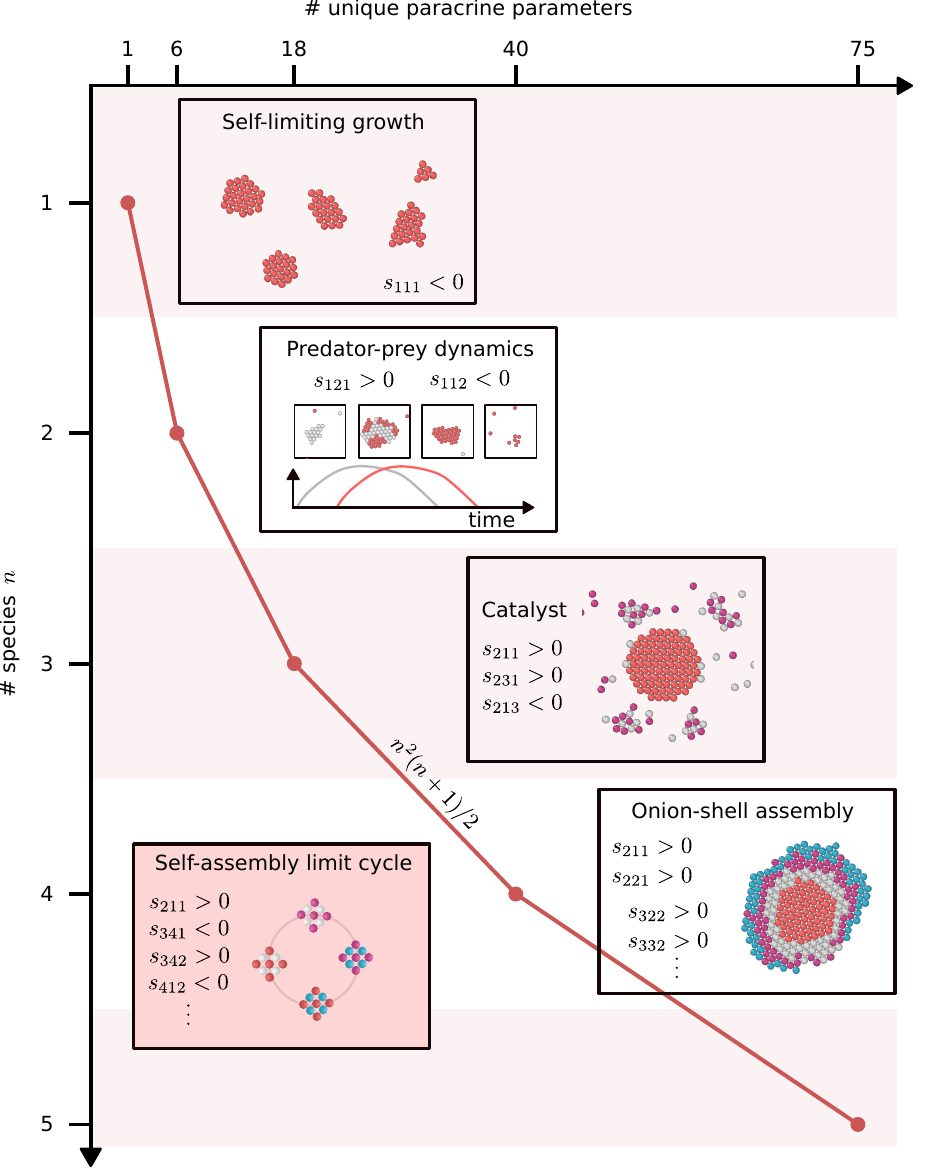}
    \caption{Number of unique elements of the rank-3 signaling matrix $s_{\alpha\beta\gamma}$ as a function of the number of species $n$, under the symmetry constraint that $s_{\alpha\beta\gamma}=s_{\beta\alpha\gamma}$. Insets show examples of systems that can be designed with different signaling matrices. (i) Self-limiting growth can be induced for one species ($n=1$), if that species inhibits the bond between all particles of this species. (ii) For $n=2$, predator-prey dynamics is obtained if species 1 promotes binding to species 2, and species 2 inhibits the binding among particles of species 1. (iii) For $n=3$, heterogeneous catalysis of 2-3 clusters can be induced when species 1 induces binding of species 2 and 3. (iv) For $n=4$, apart from the self-assembly limit cycle discussed in detail in this study, an onion-shell structure can be assembled when every species promotes binding of the next layer.}
    \label{fig:cartoonothersystems}
\end{figure*}

\end{document}

\title{Supplementary information to: Synthetic paracrine signaling of colloids drives self-assembly limit cycles 
}
\author{Tim E. Veenstra, Ren\'e van Roij, Pepijn G. Moerman, Marjolein Dijkstra}

\date{\today}
\maketitle
\section{Design space}

Within our framework of synthetic paracrine signaling, we describe the influence of signals produced by species $\gamma$ on the interaction of particles of species $\alpha$ and $\beta$ with a rank-three matrix $s_{\alpha\beta\gamma}$. To induce self-assembly cycles, we consider one specific choice of this matrix, given by Eq.~(3) of the main text and described in the Methods Section III, which contains 16 non-zero elements. However, a system with four paracrine-signaling species exhibits a much larger design space with many additional tunable parameters. Given the symmetry of the $s$-matrix with respect to interchanging the first and second index, the number of independent parameters $p$ scales as
\begin{equation}
    p(n) = \frac{n^2(n+1)}{2}, 
\end{equation}
where $n$ is the number of species. For a system containing only one species, there is only one tunable signaling parameter ($p=1$), namely  the signal produced by species $1$ that influences  binding of species $1$ with itself. Already for two species, the number of independent signaling parameters increases to $p=6$; for three species we have $p=18$, and  for four species it reaches $p=40$, etc.  The resulting design space is therefore vast, offering access to a wide variety of qualitatively distinct signaling behaviors and self-assembly pathways. To illustrate this richness, we have performed simulations for systems with varying numbers of species and different choices of the signaling matrices, as shown in Extended Figure 1. %Fig.~\ref{fig:cartoonothersystems}. 

A comprehensive exploration of this design space is beyond the scope of this paper. Nevertheless, to illustrate its versatility, we  designed four qualitatively different behaviors. For a single species, we obtain self-limiting growth. The particles are intrinsically attractive, but simultaneously  produce signals that inhibit their mutual binding.  As a  result, cluster growth is self-limited, with  a critical cluster size beyond which no further growth is possible.  For two species, we design predator-prey-like dynamics. Clusters of species 1  nucleate first and  produce signals that promote the binding of species 2 to both species 1 and 2. In turn, species 2 strongly inhibits the binding of species 1 to itself and to species 2, causing species 1 to dissolve from  the structure. The resulting structure, which is rich in species 2,  eventually disassembles due to the lack of active promotion in the absence of species 1. For three species, we design a catalytic assembly process of clusters with species 2 and 3. Here, species 1 promotes the binding of species 2 to species 1, while simultaneously promoting the binding of species 3 to species 2. However, species 3 strongly inhibits the binding between species 1 and 2, thereby ``releasing'' them from the cluster surface. This leads to the catalytic formation of clusters with species 2 and 3, even though these two species do not have intrinsic attraction. Finally, we  design the self-assembly of onion-shell structures, in which each species promotes the binding of the species forming the next layer both onto the existing  layer and among themselves.

\section{System without signal inhibition}
\begin{figure*}[]
    \centering
    \includegraphics[width=\linewidth]{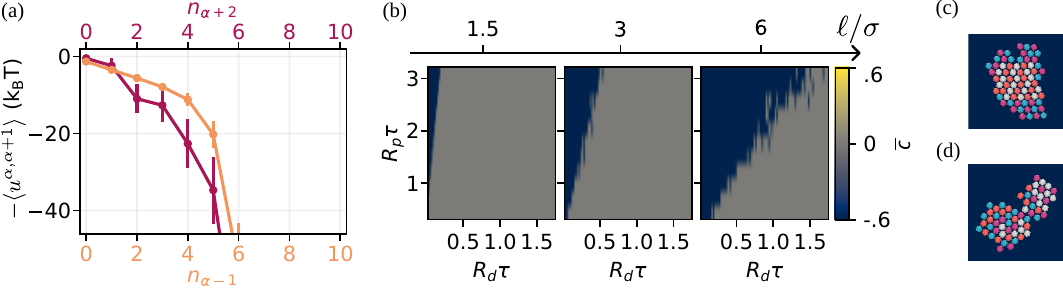}
    \caption{(a) Average interaction strength $-\langle u^{\alpha,\alpha+1}\rangle$ as a function of (red) the number of neighbors $n_{\alpha+2}$ of species $n+2$, and (orange) the number of neighbors $n_{\alpha-1}$ of species $\alpha-1$ for a modified system in which  particles can only promote interactions, while inhibition is only mediated globally through signal degradation. Data was obtained at signal production rate $R_p\tau = 3.08$, degradation rate $R_d\tau=0.22$ and signal range $\ell/\sigma=6$. Error bars represent the standard deviation of the mean. (b)   The  anti-correlation parameter $\bar{c}$, as defined in the Methods section, in the uninhibited system as a function of production rate $R_p\tau$ and degradation rate $R_d\tau$. (c-d) Typical configurations  in the uninhibited system at $R_p\tau = 3.08$, $R_d\tau=0.22$, and (c) $\ell/\sigma=6$ and (d) $\ell/\sigma=3$. 
    }
    \label{fig:uninhibited}
\end{figure*}
The emergence of the cyclic behavior as described in the main text is determined by the signaling policy defined in Eq. (3) of the main text, which combines local signal-mediated promotion ($s_{\alpha\beta\gamma}>0$) of specific interactions with inhibition ($s_{\alpha\beta\gamma}<0$) of others.  At first sight, however, it may appear that explicit inhibition is not strictly necessary for a minimal model of the emergence of stable cyclic dynamics and time-reversal symmetry breaking. The reason is that the global signal degradation rate $R_d$ used in Eq.(2) may be sufficiently strong to weaken interactions that were promoted in the past.  While a system without active inhibition reduces the ability to regulate its composition, it offers a potential advantage in implementation, since the production of only one signal per species is required instead of two.   
However, here we find that interaction-inhibiting signals are crucial as their  omission effectively eliminates cyclic behavior. To demonstrate this, we perform simulations using the dynamics of Eq.(2) of the main text with the modified signal-production matrix 
\begin{equation}
    s_{\alpha\beta\gamma} =  \delta_{\alpha, \beta-1}  \delta_{\alpha, \gamma+1} +  \delta_{\alpha,\beta+1}  \delta_{\alpha, \gamma+2},
    \label{eq:policy_uninhibited}
\end{equation}
which is element-wise non-negative. Thus, signals can only promote interactions, while inhibition arises solely from global signal degradation.   In Supplementary Fig. \ref{fig:uninhibited}(a), we show the dependence of the average interaction strength $-\langle u^{\alpha,\alpha+1}\rangle$ in the absence of inhibitory signals on the number of neighbors $n_{\alpha-1}$ (orange) and $n_{\alpha+2}$ (red), using  otherwise identical system parameters as for the explicitly inhibiting system in Fig. 4 of the main text. In contrast to the fully inhibiting system, we observe essentially no difference between the dependencies on $n_{\alpha-1}$ and $n_{\alpha+2}$ in Supplementary Fig. \ref{fig:uninhibited}(a). Instead, increasing either $n_{\alpha-1}$ or $n_{\alpha+2}$ leads to an increase in the average interaction strength  $-\langle u^{\alpha, \alpha+1} \rangle$.
This indicates that, without interaction-inhibiting signals, the paracrine-like interactions become essentially symmetric, suggesting that sustained self-assembly cycles cannot occur in the absence of inhibitory signals.  To further investigate the dynamics in the absence of inhibiting signals, we perform a parameter sweep over the  signal production rate $R_p\tau$ and degradation rate $R_d\tau$, analogous to Fig. 3(b) of the main text. We calculate the average  anti-correlation parameter $\bar{c}$ as defined in the Methods section, as shown in Supplementary Fig. \ref{fig:uninhibited}(b) for three values of the signal radius $\ell$. We observe a direct transition between the aggregation regime III and the melting regime II, without an intermediate regime of sustained self-assembly limit cycles for all values $\ell$. These findings demonstrate that explicit  inhibition of interactions is required for  time-reversal symmetry breaking and sustained self-assembly cycles. 

Compared to the system with both promotion and inhibition, the uninhibited system requires a higher signal degradation rate to melt the self-assembled aggregates. In the absence of inhibitory signals, bond destabilization is driven solely by signal degradation, making the aggregates more resistant to melting. Interestingly, the slope of the transition between aggregation and melting in the $R_p\tau$-$R_d\tau$ representation depends now strongly on $\ell$, in contrast to the system with inhibitory signaling. This dependence is absent in the presence of inhibitory signals, as the inhibitory and promoting signals act over the same length scale, thereby compensating for changes in $\ell$.

The absence of a sustained limit-cycle phase in the uninhibited system can be rationalized by considering  clusters that contain an interface between two distinct phases. We find that such  configurations are  stable, because the interaction strengths in both phases continue to increase until a steady state is reached between linear signal production and signal degradation. For the system shown in Supplementary Fig. \ref{fig:uninhibited}(a), this balance corresponds to interaction energies exceeding $100 {~k_B T}$ for particles with large numbers of neighbors. Direct inspection of simulation snapshots, shown in Supplementary Fig. \ref{fig:uninhibited}(c-d), for both $\ell/\sigma=6$ and $\ell/\sigma=3$, confirms the existence of these stable  interfaces. As expected, the characteristic length scale over which two neighboring phases can stabilize one another is set by the signal range  $\ell$. The coexistence of two opposing phases prevents the system from progressing through the sequence of distinct configurations, thereby suppressing self-assembly cycles.  
We therefore conclude that a signaling policy that includes both promotion and inhibition, as defined by the positive and negative terms of Eq. (3), constitutes the minimal model required to realize self-assembly limit cycles.

We can further rationalize the stability of the interface of two opposite phases. For the uninhibited system defined by Eq. (\ref{eq:policy_uninhibited}), the matrix representation is given by
\begin{widetext}
\begin{align*}
    s_{\alpha\beta 1} &= \left(
    \begin{matrix} 
        0 & 0 & 0 & 0 \\
        0 & 0 & 1 & 0 \\
        0 & 1 & 0 & 0 \\
        0 & 0 & 0 & 0 \\
    \end{matrix}\right)_{\alpha\beta} && 
    s_{\alpha\beta 2} = \left(
    \begin{matrix} 
        0 & 0 & 0 & 0 \\
        0 & 0 & 0 & 0 \\
        0 & 0 & 0 & 1 \\
        0 & 0 & 1 & 0 \\
    \end{matrix}\right)_{\alpha\beta} && 
    s_{\alpha\beta 3} = \left(
    \begin{matrix} 
        0 & 0 & 0 & 1 \\
        0 & 0 & 0 & 0 \\
        0 & 0 & 0 & 0 \\
        1 & 0 & 0 & 0 \\
    \end{matrix}\right)_{\alpha\beta} && 
    s_{\alpha\beta 4} = \left(
    \begin{matrix} 
        0 & 1 & 0 & 0 \\
        1 & 0 & 0 & 0 \\
        0 & 0 & 0 & 0 \\
        0 & 0 & 0 & 0 \\
    \end{matrix}\right)_{\alpha\beta} .
\end{align*}
\end{widetext}
The sum of two consecutive matrices still promotes the formation of the next structure in the cycle. However,   it is also immediately clear that  aggregates containing all species are promoted in this case; a superposition of all matrices promotes all possible bonds. As a consequence, aggregates containing all species are stabilized rather than suppressed.

\section{Signal concentration profiles}
In this work, we have approximated the signal concentration profile around a colloidal particle with an instantaneous uniform distribution of radius $\ell$. In experimental systems, however, the distribution is non-uniform and depends on the signal degradation rate in the surrounding fluid, and the signal diffusion constant. Concentration profiles $c(r)$ around signal-producing particles in the presence of signal-degradation have been derived analytically \cite{Kim2026}, and are of the form 
\begin{equation}
    c(r) \propto \frac{\ell}{r} e^{-r/\ell}
\end{equation}
To investigate the robustness of cyclic self-assembly in the presence of chemical signaling, we perform simulations with this effective signal concentration distribution. The equation of motion for the interaction strength between particles $i$ and $j$, of species $\alpha$ and $\beta$ respectively are then given by

\begin{align}
    \label{uij_2}
    \frac{\partial u^{\alpha\beta}_{ij}}{\partial t} =& -R_d u^{\alpha\beta}_{ij} + \\  & \hspace{-10mm} R_p k_BT\sum_{\gamma=1}^{n}\sum_{k=1}^{N_\gamma} \frac{1}{2}\Big(H(\ell-r_{ik})\frac{e^{-r_{ik}/\ell}}{r_{ik}/\ell} \notag\\ & \hspace{15mm} +H(\ell-r_{jk})\frac{e^{-r_{jk}/\ell}}{r_{jk}/\ell}\Big)s_{\alpha\beta\gamma}. \notag 
\end{align}

Due to the degradation, the signal is effectively screened at larger distances. We observe robust self-assembly limit cycles, even with the more realistic signal concentration profiles, as shown in Supplementary Fig. \ref{fig:kymograph_screenedconcentrations}. This simulation was performed at production rate $R_p\tau = 1.54$ and degradation rate $R_d\tau = 0.044$, with a signal radius $\ell/\sigma=12$. Slightly higher values of $\ell$ are used due to the screening effect.
\begin{figure}
    \centering
    \includegraphics[width=\linewidth]{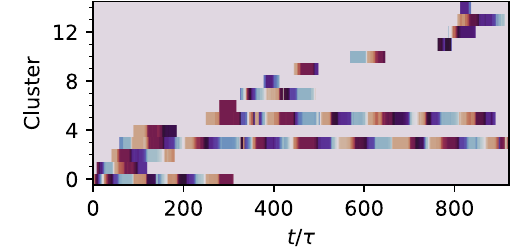}
    \caption{Kymographs of the phase $\theta_i$ of each cluster $i$ in the system as a function of time $t/\tau$ for  signal production rate $R_p\tau = 1.54$ and degradation rate $R_d\tau = 0.044$, with a signal radius $\ell/\sigma=12$}
    \label{fig:kymograph_screenedconcentrations}
\end{figure}

\section{Theoretical model of coupled Kuramoto oscillators}

All clusters in our system are characterized by an oscillatory phase angle $\theta_i$ and exchange particles with the surrounding fluid through  the absorption and release of free colloids. This process effectively couples the  clusters through the shared fluid reservoir, giving rise to the phase correlations and anti-correlations described in the main text. To capture this coupling, we introduce a simple second-order Kuramoto-like model \cite{Winfree1967, Kuramoto1975, Strogatz2000} for the phase angles $\theta_i$ of only three clusters that are coupled through the depletion of monomers in the reservoir. Although our simulations typically contain more than three clusters  at the same time,  we restrict our analysis to  the simplest non-trivial case of  three coupled oscillators, described by
\begin{equation}
    \frac{\partial \theta_i}{\partial t} = \omega  + \sum_{j=1}^3 \big( K_1 \sin{(\theta_i-\theta_j) + K_2 \sin{2(\theta_i - \theta_j)}} \big).
    \label{eomkuramoto}
\end{equation}
Here, $\omega$ is the natural frequency of the oscillators, which we consider to be constant and identical for all oscillators for the remainder of this work, and $K_1$ and $K_2$ are the coupling constants of the first and second harmonics, respectively. Intuitively, this model describes the oscillations of three clusters, where the first-harmonic coupling term $K_1 \sin{(\theta_i-\theta_j)}$  describes the effect of colloid depletion induced by cluster $j$, which slows down the evolution of cluster $i$ when $j$ is  ahead of $i$ in the cycle. In reality,  particle species are discrete, a feature which is not captured by a purely sinusoidal first-harmonic term. To describe some of additional effects arising from the discreteness of particle species, we include a second-harmonic term. These three coupled equations are most conveniently expressed in terms of phase differences  between  oscillators.  We 
therefore define $\psi = \theta_2 - \theta_1$ and $\phi=\theta_3-\theta_1$, yielding
\begin{align}
    \frac{\partial \theta_1}{\partial t} = \omega &- K_1 (\sin{(\psi)} + \sin{(\phi)}) \\  &- K_2 (\sin{(2\psi)} + \sin{(2\phi})) \notag,
\end{align}
\begin{align}
    \label{eq:psi}
    \frac{\partial \psi}{\partial t} = K_1( 2 \sin{(\psi)} + \sin{(\psi-\phi)} + \sin{(\phi)} ) \hspace{2.5em} \\ 
    +K_2 (2 \sin{(2\psi)} + \sin{(2(\psi-\phi))} + \sin{(2\phi)}) \notag,
\end{align}
\begin{align}
\label{eq:phi}
    \frac{\partial \phi}{\partial t} = K_1( 2 \sin{(\phi)} + \sin{(\phi-\psi)} + \sin{(\psi)} ) \hspace{2.5em} \\ 
    +K_2 (2 \sin{(2\phi)} + \sin{(2(\phi-\psi))} + \sin{(2\psi)}) \notag.
\end{align}
This transformation completely decouples $\psi$ and $\phi$ from $\theta_1$, reducing the system to a two-dimensional dynamic system  with time-dependent phase differences relative to the (still evolving) first oscillator. 
\begin{figure*}
    \centering
    \includegraphics[]{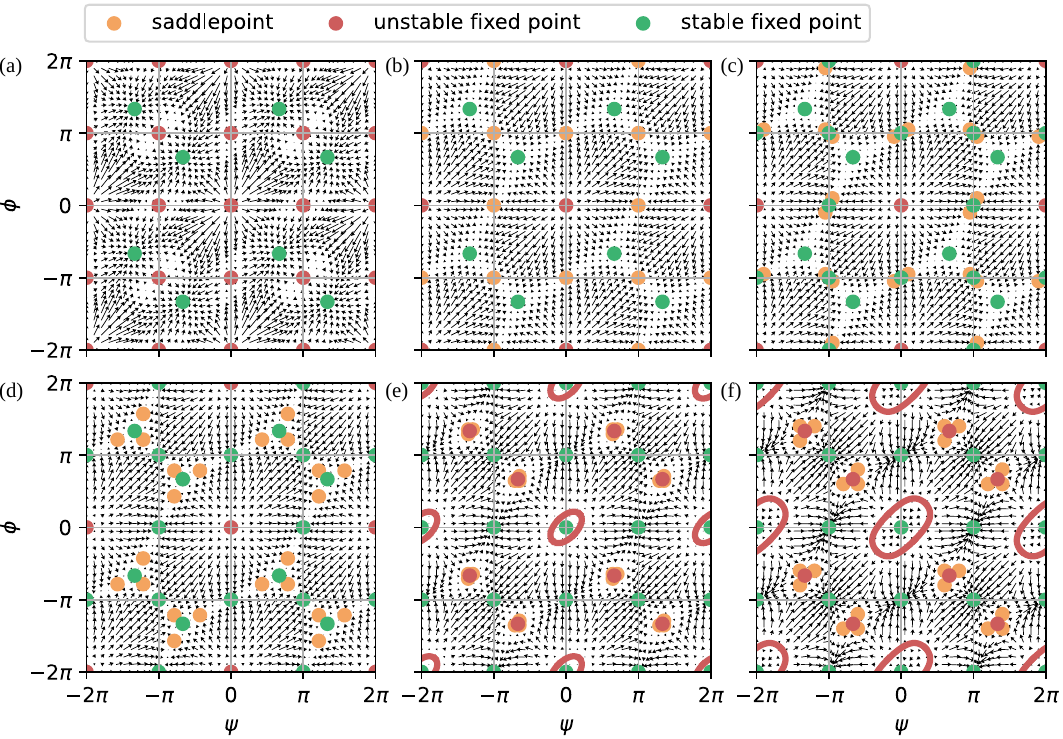}
    \caption{Vector fields $(\partial_t \psi, \partial_t \phi)$ as a function of phase differences $\psi$ and $\phi$. The points and lines correspond to fixed points of various stability. These systems were plotted with parameter values of $K_1=0.2$ and (a) $K_2=0.12$, such that $K_2>K_1/2$, (b) $K_2 = 0$, such that $-K_1/6<K_2<K_1/2$, (c) $K_2 = -0.034$, such that $K_1+6K_2 = -4\cdot 10^{-3}<0$, (d) $K_2 = -0.05$, such that $K_1+6K_2 = -10^{-1}<0$, (e) $K_2 = -0.12$, such that $K_1+2K_2=-0.04<0$, and (f) $K_2=-0.2$ such that $K_1+2K_2=-0.2<0$,}
    \label{fig:vectorfields}
\end{figure*}\\

\subsection{Fixed points and bifurcations}
This coupled two-dimensional system has several fixed points, defined by  $\partial_t \psi = \partial_t \phi = 0$. We note that the system is symmetric under the exchange $\psi \leftrightarrow \phi$, and therefore any fixed point has a corresponding symmetry-related counterpart. The trivial fixed point is located at $(\psi, \phi) = (0, 0)$, which we refer to as  the synchronization fixed point, as all oscillators are phase-locked here. Another fixed point is located at $(\psi, \phi) = (0, \pi)$, which is called the $\pi$-phase fixed point, as the oscillators distribute themselves into two clusters separated by a phase difference of  $\pi$. The splay fixed point is given by  $(\psi, \phi) = \left(2\pi/3, 4\pi/3 \right)$, in which the three oscillators are evenly  distributed  along the unit circle. For certain values of $K_1$ and $K_2$ there are additional fixed points, which will be discussed  together with the associated bifurcations below. To determine the stability of  the fixed points, we  linearize the equations around a fixed point $(\psi^*, \phi^*)$ by introducing perturbations $u=\psi-\psi^*$ and $v = \phi - \phi^*$, yielding
\begin{widetext}
\begin{align}
\label{eq:linearu}
    \partial_t u = &\left( K_1 ( 2\cos{\psi} + \cos{(\psi-\phi)} ) + K_2 (4 \cos2\psi + 2 \cos (2\psi-2\phi))\right)u \\
    &+ \left(K_1 (\cos\phi - \cos(\psi-\phi)) + K_2(2\cos2\phi - 2\cos(2\psi-2\phi) \right) v \notag,
\end{align}

\begin{align}
\label{eq:linearv}
    \partial_t v = & \left(K_1 (\cos\psi - \cos(\phi-\psi)) + K_2(2\cos2\psi - 2\cos(2\phi-2\psi) \right) u \\
    &+  \left( K_1 ( 2\cos{\phi} + \cos{(\phi-\psi)} ) + K_2 (4 \cos2\phi + 2 \cos (2\phi-2\psi))\right)v \notag, 
\end{align}
\end{widetext}
which can  be written in matrix form as a linear system of equations  
\begin{equation}
    \partial_t \begin{pmatrix} u\\v\end{pmatrix} = A \begin{pmatrix} u\\v\end{pmatrix},
\end{equation}
where $A$ is the matrix defined by Eqs.  (\ref{eq:linearu}) and (\ref{eq:linearv}), evaluated at the fixed point $(\psi^*, \phi^*)$.

\subsubsection{Synchronization fixed point}
The linearized equations around the synchronization fixed points $(0,0)$ yield the matrix 
\begin{equation}
    A = \begin{pmatrix} 3K_1 + 6 K_2 & 0\\ 0 & 3K_1 +6K_2\end{pmatrix}.
\end{equation}
This matrix has two degenerate eigenvalues $\lambda_{1,2} = 3K_1 + 6 K_2$, with corresponding  independent eigenvectors $(1,0)$ and $(0,1)$, which therefore span the plane. The fixed point is therefore a star node, since all vectors are a linear combination of the eigenvectors and get uniformly stretched by the same factor $\lambda_{1,2}$. A bifurcation occurs when the eigenvalues change sign, i.e.  at $K_1+2K_2 = 0$. The synchronization point is stable for $K_1+2K_2 < 0$ and unstable for $K_1+2K_2>0$.

\subsubsection{Splay fixed points}
The linearized equations around the splay fixed points, such as  $(2\pi/3, 4\pi/3)$, yield the matrix
\begin{equation}
    A = \begin{pmatrix} -3K_1/2 - 3 K_2 & 0\\ 0 & -3K_1/2 - 3 K_2\end{pmatrix}.
\end{equation}
This matrix also has two degenerate eigenvalues $\lambda_{1,2} = -3K_1/2 - 3K_2$, with corresponding  independent eigenvectors $(1,0)$ and $(0,1)$, and the trace is given by $\tau=-3K_1-6K_2$. This fixed point is therefore a stable star node for $K_1+2K_2>0$, and an unstable star node when $K_1+2K_2<0$.

\subsubsection{$\pi$-phase fixed points}
The linearized equations around the $\pi$-phase fixed points $(0, \pi)$ or $(\pi, \pi)$ yield
\begin{equation}
    A = \begin{pmatrix} K_1 + 6 K_2 & 0\\ 0 & -3K_1 +6K_2\end{pmatrix}.
\end{equation}
The eigenvalues are given by $\lambda_1=3(2K_2-K_1)$ and $\lambda_2=K_1 + 6K_2$. The determinant of the matrix $\Delta=\lambda_1\lambda_2$ is negative when only one of the eigenvalues is negative, \textit{i.e.} $-K_1/6<K_2<K_1/2$. In this regime, the fixed point is  a saddlepoint. Outside this interval, both eigenvalues have the same sign and the stability of the fixed point is determined by the trace $\tau=\lambda_1+\lambda_2 = -2K_1+12K_2$. For $\tau>0$, i.e. $K_2>K_1/6$, the fixed point is an unstable node, whereas  for $\tau <0$  the fixed point is a stable node. The stability of the $\pi$-phase fixed point can therefore be summarized as in Fig. \ref{fig:pitchfork}(a).

\subsubsection{Bifurcations and additional fixed points}

There are also  additional fixed points whose existence and position depend on $K_1$ and $K_2$. The bifurcation at $K_1+6K_2=0$ associated with  the $\pi$-phase fixed points $(0,\pi)$ and $(\pi, \pi)$ is accompanied by the emergence  of two other fixed points, as can also be seen in Supplementary Fig.  \ref{fig:vectorfields}. These fixed points correspond to  saddle points that are situated along the lines connecting  the $\pi$-phase   and the splay fixed points. To characterized these solutions, we parameterize the connecting lines by  a variable $s$, for example  between $(\pi, \pi)$ and $(4\pi/3, 2\pi/3)$, giving  $\psi = \pi + \pi s/3$ and $\phi = \pi-\pi s /3$, with  $s\in(0,1)$. Inserting  this Ansatz into Eq. (\ref{eq:psi}), we  numerically solve for the fixed-point condition 
\begin{equation}
     \left( \sin \frac{2\pi s}{3} - \sin \frac{\pi s}{3} \right) + \frac{K_2}{K_1} \left( \sin \frac{4\pi s}{3} + \sin \frac{2\pi s}{3} \right) = 0.
\end{equation}

\begin{figure*}
    \centering
    \includegraphics[]{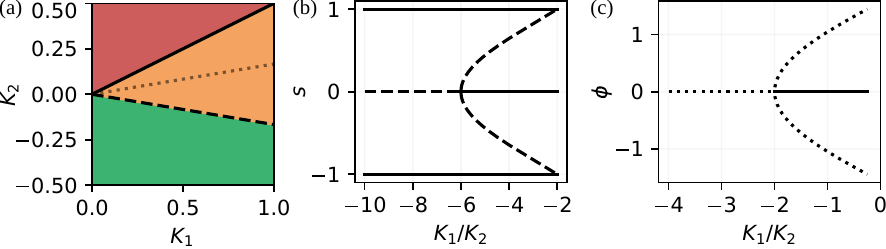}
    \caption{(a) Stability of the $\pi$-phase fixed points. The lines that demarcate the areas of instability (red), saddlepoints (orange), and stability (green) are given by $K_2=-K_1/6$ (dashed) and $K_2 = K_1/2$ (solid). (b) Bifurcation diagram of the saddlepoints between the pi-phase and splay fixed points. (c) Bifurcation diagram of the non-isolated fixed point around the synchronization fixed point. (b-c) Lines denote stable (solid), saddlepoint (dashed), and unstable (dotted) fixed points.}
    \label{fig:pitchfork}
\end{figure*}

We  plot the position $s$ along this line as a function of $K_2/K_1$ to get the bifurcation diagram as shown in Fig. \ref{fig:pitchfork}(b).

A similar situation occurs at the synchronization fixed point. At the bifurcation $K_1+2K_2=0$, an unstable,  non-isolated fixed point appears in the form of a closed curve surrounding the synchronization fixed point. To derive an analytical expression for this curve, we add Eqs. (\ref{eq:psi}) and (\ref{eq:phi}), and set this to zero. This can be written as
\begin{equation}
    \frac{K_2}{K_1} \left( \sin2\psi + \sin2\phi \right) = -(\sin\psi + \sin\phi).
\end{equation}
Using the identity $\sin\alpha+\sin\beta = 2\sin(\frac{\alpha+\beta}{2})\cos(\frac{\alpha-\beta}{2})$, and defining $\gamma\equiv\psi+\phi$ and $\delta\equiv\psi-\phi$, we find
\begin{equation}
    \frac{\sin\gamma/2}{\sin\gamma} = - \frac{K_2}{K_1} \frac{\cos\delta/2}{\cos\delta}.
\end{equation}
Since both sides of this equation are independent, we can set each side equal to a constant $\varepsilon$. We find
\begin{equation}
    \delta = \pm 4 \tan^{-1}\left(\sqrt{\frac{\sqrt{(\varepsilon K_1/K_2)^2+8}-3}{\varepsilon K_1 /K_2 -1}}\right) + 4\pi c_1 
\end{equation}
\begin{equation}
    \gamma = \pm 4 \tan^{-1}\left( \frac{\sqrt{2\varepsilon-1}}{\sqrt{2\varepsilon+1}} \right) +4\pi c_2,
\end{equation}
where $\varepsilon\in[0.5, -K_1/K_2]$ and $c_1, c_2$ are integer constants. This expression for the curve of non-isolated fixed points can be transformed back to the original coordinates, and is shown in Fig.  \ref{fig:vectorfields}(e-f). We plot the intersection of this curve with the $\psi$-axis as a function of $K_1/K_2$ in Fig.  \ref{fig:pitchfork}(c). 

\subsubsection{Oscillator trajectories}
\label{sec:phasediff}
To collect statistics on the relative phase differences of the oscillators for various values of the coupling constants, we numerically integrate the equations of motion of Eq. (\ref{eomkuramoto}). All trajectories are initialized from random points in the $(\psi, \phi)$ plane and integrated with a fourth-order Runge-Kutta scheme. The results of 500 trajectories are shown in Supplementary Figs. \ref{fig:trajectories}(a-e) for different values of $K_2$.
\begin{figure*}
    \centering
    \includegraphics[]{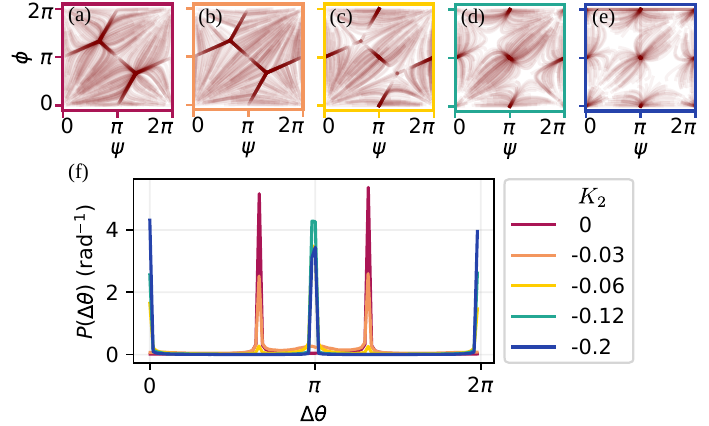}
    \caption{Oscillator trajectories integrated through time for $K_1=0.2$ and (a) $K_2 = 0$, (b) $K_2=-0.02$, (c) $K_2=-0.06$, (d) $K_2 = -0.12$, and (e) $K_2=-0.2$.  (f) The average phase difference distribution $P$ as a function of phase difference $\Delta \theta$. The box colors of (a)-(e) correspond to the line colors of (f).}
    \label{fig:trajectories}
\end{figure*}
The trajectories can be observed to accumulate at the stable fixed points. From these trajectories, we 
compute the distribution of phase differences between the oscillators,  shown in Supplementary Fig. \ref{fig:trajectories}(f) for different values of $K_2$. As expected, the positions of the peaks in the phase-difference distributions shift  from a splayed state with peaks at $\Delta \theta = 2\pi/3$, to a $\pi$-phase state with peaks at $0$ and $\pi$ upon increasing $K_2$. The latter closely resembles the phase-difference distribution as obtained from simulations of the signal-producing colloidal system. 

\subsection{Physical significance}
This Kuramoto model exhibits a range of distinct oscillator phase behaviors depending on the values of $K_1$ and $K_2$. The Kuramoto model is one of the most well-known models to describe synchronization between coupled oscillatory systems such as coupled pendulums, Josephson junctions arrays \cite{Wiesenfeld1996}, and flashing fireflies \cite{Buck1976}.  Most studies focus on $K_1<0$ and $K_2=0$, where  a large number of oscillators spontaneously synchronizes, even if they have small differences in their  natural frequencies $\omega$ \cite{Kuramoto1975}. Here, we are interested in the opposite case, where a small number of oscillators are effectively hindered by particle depletion induced by  oscillators with a positive relative phase difference. As discussed in the main text,  our simulations on signaling colloids exhibit an unexpected symmetry breaking in which  clusters separate into two groups that are synchronized but with a phase difference of exactly $\pi$ between them. This behavior is not  expected from symmetry arguments alone. Remarkably, our Kuramoto model reproduces the observed anti-phase synchronization when the second-harmonic coupling strength $K_2$ is sufficiently large. 

A direct mapping between these theoretical coupling parameters $K_1$ and $K_2$ and the control parameters in  simulations or experiments is highly challenging. Nevertheless, it is reasonable to expect that $|K_1|, |K_2| \leq 1$ , since particle depletion cannot exceed $100\%$. By comparing the relative phase-difference distributions presented in Subsection \ref{sec:phasediff}, with those obtained from our simulation on signaling colloids in  Fig. 5(c), we find  that our simulated system most closely resembles the regime characterized by  $K_2 = -0.12$, corresponding to $K_2/K_1 = -0.6$

\bibliography{library}